\documentclass{article}

\usepackage{epsfig}
\usepackage{amssymb}

\begin{document}
\null
\hfill WUE-ITP-2002-022

\hfill hep-ph/0209108\\
\vskip .4cm
\begin{center}
{\Large \bf Chargino Production and Decay in \\[.4em]
Photon-Photon-Collisions}\\[.4em]
\vskip 1.5em

{\large
{\sc
T.~Mayer,
C.~Bl\"ochinger\footnote{e-mail:
bloechi@physik.uni-wuerzburg.de},
F.~Franke\footnote{e-mail:
fabian@physik.uni-wuerzburg.de},
H.~Fraas\footnote{e-mail:
fraas@physik.uni-wuerzburg.de}
}}\\[3ex] 
{\footnotesize \it
Institut f\"ur Theoretische Physik und Astrophysik, Universit\"at
W\"urzburg, \\D--97074 W\"urzburg, Germany}
\vskip .5em
\par
\vskip .4cm
\end{center}

\begin{abstract}
  
We discuss the pair production of charginos in collisions of polarized photons
$\gamma\gamma\rightarrow\tilde{\chi}_i^+\tilde{\chi}_i^-$, ($i=1,2$)
and the subsequent leptonic decay of the lighter chargino
$\tilde{\chi}_1^+\rightarrow\tilde{\chi}_1^0e^+\nu_e$ including the
complete spin correlations. Analytical formulae are given for the
polarization and the spin-spin correlations of the charginos. Since
the production is a pure QED process  the
decay dynamics can be studied separately. For high energy photons
from Compton backscattering of polarized laser pulses off polarized
electron beams numerical results are presented for the cross section,
the angular distribution and the forward-backward asymmetry of the
decay positron. Finally we study  the dependence on the gaugino mass parameter
$M_1$ and on the sneutrino mass for a gaugino-like
MSSM scenario.

\end{abstract}

\section{Introduction}

From both a theoretical and a phenomenological point of view supersymmetry
(SUSY) is the most attractive concept for new physics beyond the
Standard Model. The first candidates for supersymmetric particles
are expected to be discovered at the LHC. Then precision measurements are
necessary  to identify the specific supersymmetric scenario which is
realized in nature and to determine the model parameters. Due to the
clear signatures and the feasibility of polarized beams
Linear Colliders offer outstanding possibilites to
find SUSY particles and to study their properties \cite{tdr3}.
Here charginos, the supersymmetric partners of the charged gauge
and Higgs bosons, are of particular interest as they are expected to
be light enough to be produced with comparably large cross
sections at an $e^+e^-$ collider. In the Minimal Supersymmetric Standard Model 
(MSSM) the production process is determined by 
the SUSY parameters $M_2$, $\mu$, $\tan\beta$ and 
the sneutrino mass $m_{\tilde{\nu}_e}$, whereas the chargino
decay into the 
lightest neutralino $\tilde{\chi}^0_1$ which we assume to be the 
lightest supersymmetric particle (LSP)
depends in addition on the 
gaugino mass parameter $M_1$ and the mass $m_{\tilde{e}_L}$ 
of the left selectron.
The SUSY parameters can be determined with high precision in a
combined analysis of neutralino and chargino pair production at an 
$e^+e^-$ collider with polarized beams \cite{choi}.
 
Besides for the $e^+e^-$ option many studies for the $\gamma\gamma$ mode
of a Linear Collider have been performed with high luminosity polarized
photon beams obtained by Compton-backscattering of laser pulses off the
electron beams \cite{tdr6}. In the present paper we study chargino pair 
production
$\gamma\gamma\rightarrow\tilde{\chi}^+_i\tilde{\chi}^-_i (i=1,2)$ in photon
collisions and the subsequent leptonic chargino decay.
In our numerical analysis we focus on the decay of the light
positive chargino 
$\tilde{\chi}^+_1\rightarrow\tilde{\chi}^0_1e^+\nu_e$, but 
consider in the analytical formulae the complete spin correlations
between the production and decay process.

Since the production is a pure QED process at tree level 
which depends only on the 
mass of the charginos
the chargino decay mechanism can be studied separately 
in $\gamma\gamma$ collisions.
Provided the chargino mass has been measured and 
the energy spectrum and polarization of the high energy photons are well under 
control the production cross section and the polarization of the
charginos are 
known and can be varied by suitable choice of the polarization of the 
laser photons and the converted electron beam. 
Then the direct measurement of the chargino decay 
branching ratios for the various decay channels may be useful for
an analysis of the chargino system
complementary to the $e^+e^-$ mode where both production and decay
are sensitive to the SUSY parameters.

This paper is organized as follows:
In Sect.~2 we present the general spin density matrix formalism for the production of 
fermions with polarized photon beams and their subsequent decay. 
In order to analyze the influence of the fermion polarization on the forward-backward
asymmetry of the decay product one needs the 
spin density production matrix which is given analytically in the $\gamma\gamma$-CMS
for circularly polarized photons in Sect.~3. 
For two representative SUSY scenarios we present in Sect.~4 
numerical results for the chargino production cross section,
the branching ratio of the leptonic decay
of the lighter chargino $\tilde{\chi}^+_1$, 
the angular distribution and the forward-backward
asymmetry of the decay positron in the laboratory system (ee-CMS).
Finally we study the dependence on the gaugino
mass parameter $M_1$ and on the sneutrino mass in a scenario with a gaugino-like
LSP. 

\section{Helicity amplitudes and cross sections}

The analytical formulae for the differential
cross section of the combined process of chargino production by collisions of polarized photon beams 
\begin{equation}\label{produk}
\gamma_1(k)+\gamma_2(k')\to \tilde{\chi}^-_k(p) + \tilde{\chi}^+_k(p') 
\quad(k=1,2)
\end{equation}
and the leptonic
decays
\begin{eqnarray}\label{zerf1}
\tilde{\chi}^-_k(p)&\to&\tilde{\chi}^0_1(q_4)+e^-(q_5)+\bar{\nu}_e(q_6) \\
\tilde{\chi}^+_k(p')&\to&\tilde{\chi}^0_1(q_1)+e^+(q_2)+\nu_e(q_3)
\label{zerf2} 
\end{eqnarray}
with complete spin correlations are calculated using the same formalism as in
\cite{gudichar} for $e^+e^-$-annihilation.

\begin{figure}[p]
\centering
\setlength{\unitlength}{1cm}

\begin{picture}(14,8.)

\put(-3.8,-19.){\includegraphics{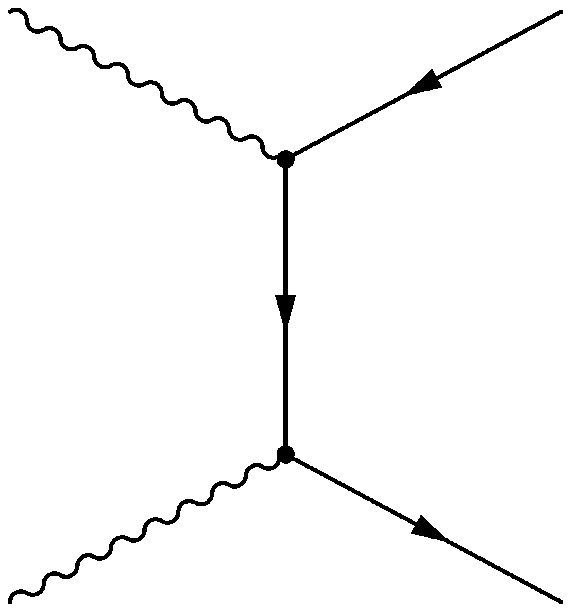}}
\put(3.3,-19.){\includegraphics{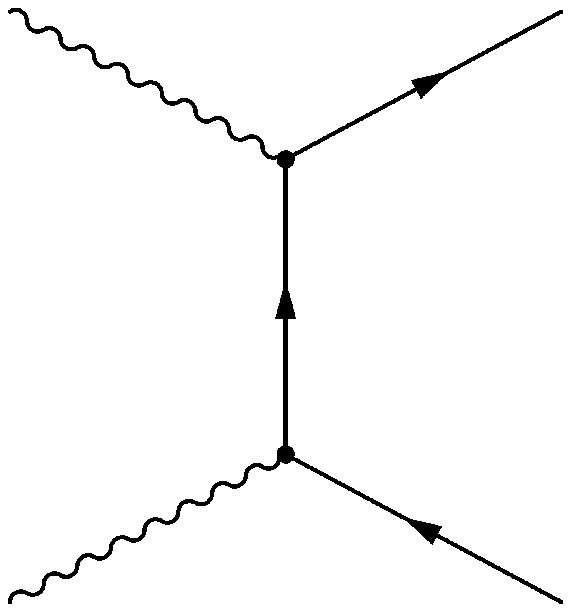}}
\put(0.2,0.4){$\gamma_2(k',\varepsilon'^{(\alpha)})$}
\put(0.2,7.13){$\gamma_1(k,\varepsilon^{(\beta)})$}
\put(5.1,0.4){$\tilde{\chi}^+_1(p',s'^b)$}
\put(5.1,7.13){$\tilde{\chi}^-_1(p,s^a)$}
\put(3.6,3.8){$\tilde{\chi}^\pm_1$}
\put(7.3,7.13){$\gamma_1(k,\varepsilon^{(\beta)})$}
\put(7.3,0.4){$\gamma_2(k',\varepsilon'^{(\alpha)})$}
\put(10.7,3.8){$\tilde{\chi}^\pm_1$}
\put(12.3,7.13){$\tilde{\chi}^+_1(p',s'^b)$}
\put(12.3,0.4){$\tilde{\chi}^-_1(p,s^a)$}

\end{picture}
\caption[]{\label{feyngraph}Feynman diagrams for the process $\gamma_1(k,\varepsilon^{(\beta)})+\gamma_2(k',\varepsilon'^{(\alpha)})\to
\tilde{\chi}^+_1(p',s'^b)+\tilde{\chi}^-_1(p,s^a)$}

\end{figure}

The production (\ref{produk}) is 
a pure QED process in leading order perturbation theory and depends only on
the chargino mass $m_k$, which is expected to be measured in $e^+e^-$
annihilation \cite{tdr3}. The helicity amplitudes 
corresponding to the Feynman diagrams in Fig.~\ref{feyngraph} are
\begin{equation}\label{ampl}
T_{P,\alpha\beta}^{\lambda_i\lambda_j}=e^2\bar{u}(p',\lambda_j)\big\{\not\!{\varepsilon}^{\prime(\alpha)}\frac{\not\!{Q}+ 
  m_k}{Q^2-m_k^2}\not\!{\varepsilon}^{(\beta)}+\not\!{\varepsilon}^{(\beta)}\frac{\not\!{Q'}+ 
  m_k}{Q'^2-m_k^2}\not\!{\varepsilon}^{\prime(\alpha)} \big\}v(p,\lambda_i) 
\end{equation}
where $p$, $\lambda_i$ ($p'$, $\lambda_j$)
are the momenta and helicities of the chargino
$\tilde{\chi}^-_k$ ($\tilde{\chi}^+_k$).
The exchanged
charginos have momenta  $Q=k-p$ and $Q'=k'-p$ and the polarization
vectors of the photons with momenta $k$, $k'$ are denoted by
$\varepsilon^{(\beta)}$ and $\varepsilon^{\prime(\alpha)}$,
respectively.  
In the following we consider only right ($\alpha$, $\beta =1$) or left
($\alpha$, $\beta =-1$) circularly polarized photons. 

The amplitude for production and decay reads
\begin{equation}
T_{\alpha\beta}=\Delta(\tilde{\chi}^-_k)\Delta(\tilde{\chi}_k^+)\sum_{\lambda_i,\lambda_j}T_{P,\alpha\beta}^{\lambda_i\lambda_j}T_D^{\lambda_i}T_D^{\lambda_j}
\end{equation}
with the propagators of the charginos with mass $m_k$ and width $\Gamma_k$ 
\begin{eqnarray}
\Delta(\tilde{\chi}^-_k) & = & \frac{1}{p^2-m_k^2+im_k\Gamma_k} \\
\Delta(\tilde{\chi}^+_k) & = &\frac{1}{p'^2-m_k^2+im_k\Gamma_k} \; .
\end{eqnarray}
For these propagators we use the narrow width approximation.

\begin{figure}[p]
\centering
\setlength{\unitlength}{1cm}

\begin{picture}(14,15.8)

\put(-3.,-12.){\includegraphics{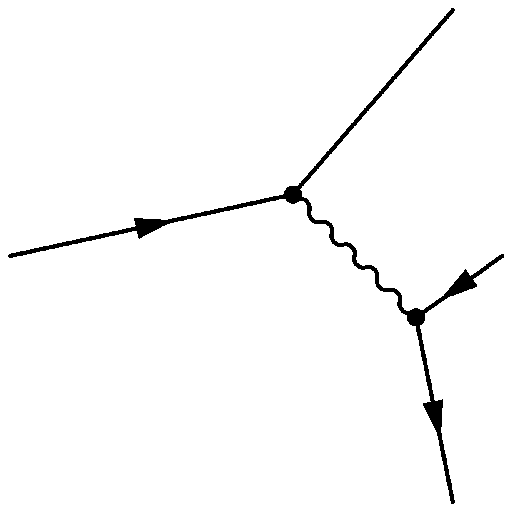}}
\put(4.,-12.){\includegraphics{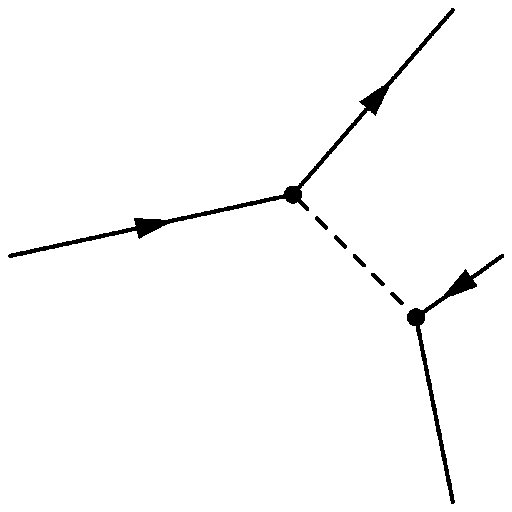}}
\put(0.5,-19.5){\includegraphics{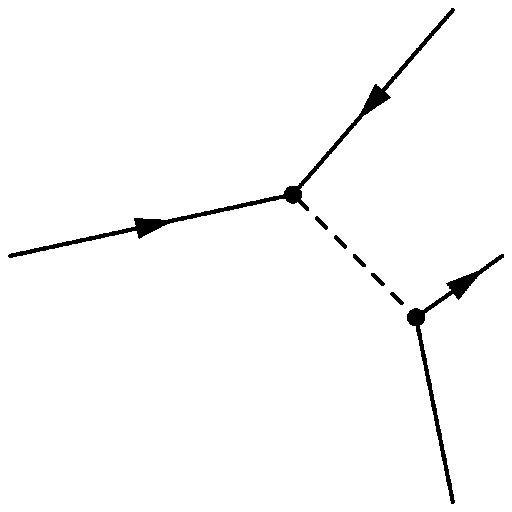}}
\put(0.5,12.){$\tilde{\chi}^+_1(p',s'^b)$}
\put(7.5,12.){$\tilde{\chi}^+_1(p',s'^b)$}
\put(4.,4.5){$\tilde{\chi}^+_1(p',s'^b)$}
\put(3.4,11.1){$W^+$}
\put(5.,14.2){$\tilde{\chi}^0_1(q_1)$}
\put(5.8,11.4){$e^+(q_2)$}
\put(4.9,8.6){$\nu_e(q_3)$}
\put(12.,14.2){$\nu_e(q_3)$}
\put(10.6,11.2){$\tilde{e}_L$}
\put(12.8,11.4){$e^+(q_2)$}
\put(11.9,8.4){$\tilde{\chi}^0_1(q_1)$}
\put(8.5,6.6){$e^+(q_2)$}
\put(7.2,3.6){$\tilde{\nu}_e$}
\put(8.3,0.9){$\tilde{\chi}^0_1(q_1)$}
\put(9.3,4){$\nu_e(q_3)$}
\put(4.7,11.9){$O^{L,R}_{11}$}
\put(4.6,12.){\vector(-1,0){0.8}}
\put(11.65,11.9){$U_{11}$}
\put(11.55,12.){\vector(-1,0){0.8}}
\put(12.9,10.6){$f^L_{e1}$}
\put(12.8,10.7){\vector(-1,0){0.8}}
\put(8.2,4.4){$V_{11}$}
\put(8.1,4.5){\vector(-1,0){0.8}}
\put(9.4,3.1){$f^L_{\nu 1}$}
\put(9.3,3.2){\vector(-1,0){0.8}}

\end{picture}
\caption[]{\label{feyngraph2}Feynman diagrams for the decay process
 $\tilde{\chi}^+_1(p',s'^b) \to
 \tilde{\chi}^0_1(q_1)+e^+(q_2)+\nu_e(q_3)$. Couplings as defined in
 \cite{gudichar}.}

\end{figure}

The helicity amplitudes $T_D^{\lambda_i}(\tilde{\chi}^-_k)$ and
$T_D^{\lambda_j}(\tilde{\chi}^+_k)$ for the leptonic decays
(\ref{zerf1}) and (\ref{zerf2}) of the charginos via $W^\pm$,
$\tilde{e}_L$ and $\tilde{\nu}_e$ exchange (Fig.\ \ref{feyngraph2})
are given in \cite{gudichar, choi1}. They depend on 
the gaugino mass parameters $M_1$, $M_2$, the higgsino mass
parameter $\mu$, the ratio $\tan\beta$ of the vacuum expectation values 
of the neutral Higgs fields
and the left selectron and the sneutrino mass.

The differential cross section for production and subsequent decay of both charginos 
in the $\gamma\gamma$-CMS is
\begin{equation}\label{cross}
d\sigma_{\alpha\beta}=\frac{1}{8E^2}|T_{\alpha\beta}|^2(2\pi)^4\delta^4(k+k'-\sum_iq_i)d\mathrm{lips}(q_1,q_2,q_3,q_4,q_5,q_6) 
\end{equation}
where $E$ is the energy of the photon beam and $d\mathrm{lips}(q_1,q_2,q_3,q_4,q_5,q_6)$ denotes the
Lorentz invariant phase space element. 

The amplitude squared (using the sum convention) 
\begin{equation} 
|T_{\alpha\beta}|^2=|\Delta(\tilde{\chi}^-_k)|^2|\Delta(\tilde{\chi}^+_k)|^2\rho^{\lambda_i\lambda_j,\lambda_i^\prime\lambda_j^\prime}_{P,\alpha\beta}\rho_D^{\lambda_i^\prime\lambda_i}(\tilde{\chi}^-_k)\rho_D^{\lambda^\prime_j\lambda_j}(\tilde{\chi}^+_k) 
\end{equation}
is composed of the unnormalized spin density
production matrix
\begin{equation}\label{mat1}
\rho^{\lambda_i\lambda_j,\lambda^\prime_i\lambda^\prime_j}_{P,\alpha\beta}=T^{\lambda_i\lambda_j}_{P,\alpha
  \beta}\;T^{\lambda^\prime_i\lambda^\prime_j\ast}_{P,\alpha \beta}
\end{equation}
and the decay matrices
\begin{equation}\label{mat2}
\rho_D^{\lambda_i^\prime\lambda_i}(\tilde{\chi}^-_k)
=T_D^{\lambda_i}T_D^{\lambda^\prime_i\ast}\quad
\rho_D^{\lambda_j^\prime\lambda_j}(\tilde{\chi}^+_k)=
T_D^{\lambda_j}T_D^{\lambda^\prime_j\ast} \; .
\end{equation}
Introducing a suitable set of polarization vectors $s^a$ ($s^{\prime
  b}$) for the chargino $\tilde{\chi}^-_k$ ($\tilde{\chi}^+_k$) one
can expand the density matrices (\ref{mat1}), (\ref{mat2}) in terms of
Pauli matrices \cite{haber} 
\begin{eqnarray}
\rho^{\lambda_i\lambda_j,\lambda^\prime_i\lambda^\prime_j}_{P,\alpha\beta}&=&\delta_{\lambda_i\lambda_i'}\delta_{\lambda_j\lambda_j'}P_{\alpha\beta}+\delta_{\lambda_j\lambda_j'}\sigma^a_{\lambda_i\lambda_i'}\Sigma^a_{P,\alpha\beta}+\delta_{\lambda_i\lambda_i'}\sigma^b_{\lambda_j\lambda_j'}\Sigma^{\prime 
  b}_{P,\alpha\beta}\\
&&+\sigma^a_{\lambda_i\lambda_i'}\sigma^b_{\lambda_j\lambda_j'}\Sigma^{ab}_{P,\alpha\beta}\\
\rho_D^{\lambda_i^\prime\lambda_i}(\tilde{\chi}^-_k)&
=&\delta_{\lambda_i^\prime\lambda_i}D_i+\sigma^a_{\lambda_i^\prime\lambda_i}\Sigma^a_D\\
\rho_D^{\lambda_j^\prime\lambda_j}(\tilde{\chi}^+_k)&
=&\delta_{\lambda_j^\prime\lambda_j}D_j+\sigma^b_{\lambda_j^\prime\lambda_j}\Sigma^{\prime b}_D
\end{eqnarray}
and obtains 
\begin{eqnarray}\label{ampli}
|T_{\alpha\beta}|^2&=&4|\Delta(\tilde{\chi}^-_k)|^2|
\Delta(\tilde{\chi}^+_k)|^2(P_{\alpha\beta}D_iD_j+D_j
\Sigma^a_{P,\alpha\beta}\Sigma^a_D\nonumber 
\\ &&\quad +D_i\Sigma^{\prime b}_{P,\alpha\beta}\Sigma^{\prime
  b}_D+\Sigma^{ab}_{P,\alpha\beta}\Sigma^a_D\Sigma^{\prime b}_D) \; .
\end{eqnarray}
The ratio
$\Sigma^a_{P,\alpha\beta}/P_{\alpha\beta}$
($\Sigma^{\prime b}_{P,\alpha\beta}/P_{\alpha\beta}$) describes the
polarization of the chargino $\tilde{\chi}^-_k$ ($\tilde{\chi}^+_k$).
$\Sigma^{ab}_{P,\alpha\beta}$ originates from spin-spin correlations
between both charginos. The analytical formulae for the quantities
$P_{\alpha\beta}$, $\Sigma^a_{P,\alpha\beta}$, $\Sigma^{\prime
  b}_{P,\alpha\beta}$ and $\Sigma^{ab}_{P,\alpha\beta}$ are given in
the next section. Analytical expressions for the decay matrices for the
leptonic decays (\ref{zerf1}) and (\ref{zerf2}) can be found in
\cite{gudichar, choi1}. 

If the decay of only one of the charginos, 
$\tilde{\chi}^+_k(p')\to\tilde{\chi}^0_1(q_1)e^+(q_2)\nu_e(q_3)$, is 
considered it is $D_i=1$, $\Sigma^a_D=0$ and
$\Delta(\tilde{\chi}^-_k)=1$ in (\ref{ampli}). Replacing the phase space element in
(\ref{cross}) by
$d\mathrm{lips}(p,q_1,q_2,q_3)$ and integrating over the phase space
of the LSP and the neutrino lead to the cross section 
$d\sigma_{e,\alpha\beta}$ of the decay positron. 
With the substitution 
$|T_{\alpha\beta}|^2=4P_{\alpha\beta}$ in (\ref{cross}) 
and the phase space element  
$d\mathrm{lips}(p,p')$ one obtains the chargino production cross section
$d\sigma_{P, \alpha\beta}$.

The optimal source of high energy polarized photon beams is Compton
backscattering of intense laser pulses off one of the beams of a
linear collider in the $e^-e^-$ mode. 
The energy distribution $P(y)$ and the mean helicity $\lambda(y)$
strongly depend on the
polarizations $\lambda_L$ of the laser photons 
and $\lambda_c$ of the converted electrons. The
analytical formulas can be found in \cite{tdr6, ginzburg}.
To obtain the chargino production cross section 
$d\sigma_{P}(s_{\gamma\gamma})/d\cos\theta$
for circularly polarized photons one has to weight the cross section   
$d\sigma_{P, \alpha\beta}(s_{\gamma\gamma}) / d\cos\theta$
with the mean helicity $\lambda(y_{1})$
and $\lambda(y_{2})$ of both beams
\begin{equation}\label{wichten}
\frac{d\sigma_p(s_{\gamma\gamma})}{d\cos\theta}=\frac{1}{4}\sum_{\alpha,\beta=\pm1}\big(1+\alpha\lambda(y_1)\big)\big(1+\beta\lambda(y_2)\big)
\frac{d\sigma_{p, \alpha\beta}(s_{\gamma\gamma})}{d\cos\theta} \; .
\end{equation}
Here $y_1=E_{\gamma_1}/E_{e_1}$
($y_2=E_{\gamma_2}/E_{e_2}$) is the ratio of the energies of
the high energy photons $\gamma_1(k)$ ($\gamma_2(k')$) and of the
energies of the respective converted electrons $e_1$
($e_2$). 
Convoluting the cross section (\ref{wichten}) with the energy distribution
$P(y_1)$ and $P(y_2)$ of the high energy photons one obtains the differential 
cross section in the laboratory frame (ee-CMS) \cite{kon}
\begin{equation}\label{conv2}
\frac{d\sigma_p(s_{ee})}{d\cos\theta^L} = \int  P(y_1)P(y_2) 
      \frac{d\sigma_p}{d\cos\theta}(\cos\theta(\cos\theta^L),s_{\gamma\gamma}=y_1y_2s_{ee})
      \frac{d\cos\theta}{d\cos\theta^L}dy_1dy_2 \; .
\end{equation}
As indicated in (\ref{conv2}) the scattering angle $\theta$ in the 
$\gamma\gamma$-CMS has to be expressed by the scattering angle $\theta^L$
in the ee-CMS:
\begin{equation}
\cos\theta=  \frac {y_2(1+\cos\theta^L)-y_1
  (1-\cos\theta^L)}{y_2(1+\cos\theta^L)+ y_1(1-\cos\theta^L)} \; . 
\end{equation}
Weighting the total cross section $d\sigma_p(s_{\gamma\gamma})$
as in (\ref{wichten}) with the mean helicities one obtains the total cross section
$d\sigma_p(s_{ee})$ in the ee-CMS by the convolution \cite{hesselb}
\begin{equation}\label{conv}
\sigma_p(s_{ee})=\int
P(y_1)P(y_2)\sigma_p(s_{\gamma\gamma}=y_1y_2s_{ee})dy_1dy_2
\end{equation}
The same procedure applies to the differential cross section 
$d\sigma_e(s_{ee})/d\cos\theta^L_{e^+}$ and the total cross section
$\sigma_e(s_{ee})$ of the positrons from chargino production and
subsequent leptonic decay.

In the laboratory system ($ee$-CMS) it is $E_1=E_2=E$ and
$s_{\gamma\gamma}=y_1y_2s_{ee}$ with $\sqrt{s_{ee}}=2E$.
To prevent $e^+e^-$ pair production by scattering of the photon beam and the laser beam
the ratio $y$ has to be adjusted to $y\leq 0.83$ which leads to
$\sqrt{s_{\gamma\gamma}^{max}}=0.83\sqrt{s_{ee}}$.

\section{The spin-density production matrix}

In this section we give the analytical formulae for the quantities
$P_{\alpha\beta}$, $\Sigma^a_{P,\alpha\beta}$, $\Sigma^{\prime
  b}_{P,\alpha\beta}$, $\Sigma^{ab}_{P,\alpha\beta}$
in (\ref{ampli}) for production of charged spin-$\frac{1}{2}$ fermions
in collisions of circularly polarized photons.
For our study of the combined process of production and decay
it is convenient to choose a coordinate frame where the
momenta are given by
\begin{eqnarray}\label{coord}
k^\mu &=&E(1,-\sin\theta,0,\cos\theta)\\
k^{\prime\mu} &=&E(1,\sin\theta,0,-\cos\theta)\\
p^\mu &=&(E,0,0,-q)\\
p^{\prime\mu} &=&(E,0,0,q)\label{coord2}
\end{eqnarray}
with $q=|\vec{p}|=|\vec{p}'|$.
For the chargino $\tilde{\chi}^-_k$ ($\tilde{\chi}^+_k$) with momentum 
$p$ ($p'$) and mass $m_k$ we introduce three space-like polarization vectors
$s^{a\mu}$ ($s^{\prime b\mu}$) ($a,b=1,2,3$), which together with
$\frac{p^\mu}{m_k}$ ($\frac{p^{\prime \mu}}{m_k}$) form an
orthonormal set
\begin{eqnarray}
s^{1\mu}&=&(0,-1,0,0)\\
s^{2\mu}&=&(0,0,1,0)\\
s^{3\mu}&=&\frac{1}{m_k}(q,0,0,-E)\\
s^{\prime 1\mu}&=&(0,1,0,0)\\
s^{\prime 2\mu}&=&(0,0,1,0)\\
s^{\prime 3\mu}&=&\frac{1}{m_k}(q,0,0,E)
\end{eqnarray}
Here $s^3$ ($s^{\prime 3}$) denotes the longitudinal polarization,
$s^1$ ($s^{\prime 1}$) the transverse polarization in the production
plane and $s^2$ ($s^{\prime 2}$) the transverse polarization
perpendicular to the production plane. With our choice of the
coordinate frame the polarization
vectors for circularly polarized photons are
\begin{eqnarray}
\varepsilon^{\prime
  (\alpha)}_{\mu}(\gamma_2)&=&\frac{1}{\sqrt{2}}(0,-\alpha\cos\theta,i,-\alpha\sin\theta)\\
\varepsilon^{(\beta)}_{\mu}(\gamma_1)&=&\frac{1}{\sqrt{2}}
(0,\beta\cos\theta,i,\beta\sin\theta)
\end{eqnarray}
with $\alpha$, $\beta=+1$ ($\alpha$, $\beta=-1$) for right (left) circularly
polarized photons.

The expression $P_{\alpha\beta}$ is independent of the chargino
polarization and reads
\begin{equation}\label{P}
P_{\alpha\beta}=C\big\{(1+\alpha\beta)m_k^2(2E^2-m_k^2)+
(1-\alpha\beta)q^2\sin^2\theta(2E^2-q^2\sin^2\theta)\big\}
\end{equation}
with 
\begin{equation}
C=\frac{e^4}{(E^2-q^2\cos^2\theta)^2} \; .
\end{equation}

The contributions $\Sigma^a_{P,\alpha\beta}$ of the
polarization of the chargino $\tilde{\chi}^-_k$ are
\begin{eqnarray}\label{pol1}
\Sigma^1_{P,\alpha\beta}&=&-2C(\alpha-\beta)Eq^2 m_k\sin^3\theta\\
\Sigma^2_{P,\alpha\beta}&=& \hspace{.28cm} 0 
\label{pol2} \\
\Sigma^3_{P,\alpha\beta}&=&
\hspace{.28cm} 2CEq\big\{(\alpha+\beta)m_k^2+(\alpha-\beta)
Eq\sin^2\theta\cos\theta\big\} \; . 
\label{pol3}
\end{eqnarray}

To obtain the contributions $\Sigma^{\prime b}_{P,\alpha\beta}$ from
the polarization of $\tilde{\chi}^+_k$ one has to exchange $\alpha$
and $\beta$ in  (\ref{pol1}) -- (\ref{pol2}).

For the quantities $\Sigma^{a b}_{P,\alpha\beta}$ describing the 
spin-spin-correlations between both charginos one obtains
\begin{eqnarray}\label{spinspin1}
\Sigma^{11}_{P,\alpha\beta}&=&
\hspace{.28cm} C\big\{(1+\alpha\beta)m_k^4-(1-\alpha\beta)q^2(E^2+m_k^2)\sin^4\theta\big\}\\ 
\Sigma^{22}_{P,\alpha\beta}&=&-C\big\{(1+\alpha\beta)m_k^4+
  (1-\alpha\beta)q^4\sin^4\theta\big\}\\ 
\Sigma^{33}_{P,\alpha\beta}&=&
\hspace{.28cm} C\big\{(1+\alpha\beta)m_k^2(2E^2-m_k^2)\nonumber\\ 
&& \hspace{.8cm} -(1-\alpha\beta)q^2\sin^2\theta
\left[2E^2-(E^2+m_k^2)\sin^2\theta\right]\big\}\\ 
\Sigma^{13}_{P,\alpha\beta}&=&
\hspace{.28cm} C(1-\alpha\beta)2Eq^2 m_k\sin^3\theta\cos\theta = 
\Sigma^{31}_{P,\alpha\beta}\\
\Sigma^{12}_{P,\alpha\beta} & = & 
\hspace{.28cm} 0 \hspace{.28cm} = \Sigma^{21}_{P,\alpha\beta}=
\Sigma^{23}_{P,\alpha\beta}=\Sigma^{32}_{P,\alpha\beta}\; .
\label{spinspin2}
\end{eqnarray} 

For unpolarized photon beams it is $\alpha=0$ or/and $\beta=0$. If both beams
are unpolarized the charginos are also
unpolarized and the spin correlations between production and decay
of one of the charginos vanish. If, however, the decay
of both charginos is considered the spin-spin correlations
(\ref{spinspin1})-(\ref{spinspin2}) are crucial for the
distribution of the opening angle between the decay products of both
of them.

From $P_{\alpha\beta}$ (\ref{P}) one obtains the differential
cross section for production of charged
spin-$\frac{1}{2}$-fermions with polarized photon beams:
\begin{equation}
\frac{d\sigma_{p,\alpha\beta}}{d\cos\theta}=\frac{q}{32\pi
  E^3}P_{\alpha\beta}
\end{equation}
and the total cross section
\begin{eqnarray}\label{prod}
\sigma_{p,\alpha\beta}&=&\frac{e^4}{16\pi
  E^6}\Bigg\{\big[m_k^2(2E^2-m_k^2)+2E^4(1-\alpha\beta)\big]\ln\frac{E+q}{m_k}
  \nonumber \\
&&+Eq\big[2E^2-m_k^2-3E^2(1-\alpha\beta)\big]\Bigg\} \; . 
\end{eqnarray}
Both the differential and the total production cross sections are sensitive 
to photon polarization only if both beams are polarized. In
addition the cross section does not change by replacing both $\alpha \to -\alpha$
and $\beta \to -\beta$.

\section{Numerical results}

In the MSSM the masses and couplings of the charginos and neutralinos
are determined by the parameters $M_1$, $M_2$, $\mu$ and $\tan\beta$
with $M_1$ usually fixed by the GUT relation
$M_1=\frac{5}{3}M_2\tan^2\theta_W$.
The parameters are chosen to be real assuming CP-conservation. 

We study pair production and leptonic decay of the lighter chargino in 
two representative scenarios A \cite{ambro} and B with the parameters given in Table
\ref{szen}. In scenario A both the lighter chargino and 
the LSP are gaugino-like while in scenario B $\tilde{\chi}^\pm_1$ and $\tilde{\chi}^0_1$ are Higgsino-like. The choice of the slepton masses corresponds to a common scalar mass
$m_0=100$ GeV. The masses of $\tilde{\nu}_e$ and
$\tilde{e}_L$ are at tree level connected by the $SU(2)_L$
relation
\begin{equation}\label{su2}
m^2_{\tilde{e}_L}=m^2_{\tilde{\nu}_e}-m^2_{W}\cos2\beta \;.
\end{equation}
 In sections \ref{secm1} and \ref{secsneu} we 
study the dependence of the cross section
$\sigma_e$ and of the forward-backward asymmetry on $M_1$ and 
$m_{\tilde{\nu}_e}$.

\subsection{Cross sections for chargino production}

\begin{figure}[p]
\centering
\setlength{\unitlength}{1cm}

\begin{picture}(11.7,6.7)

\put(0.,-2.){\includegraphics{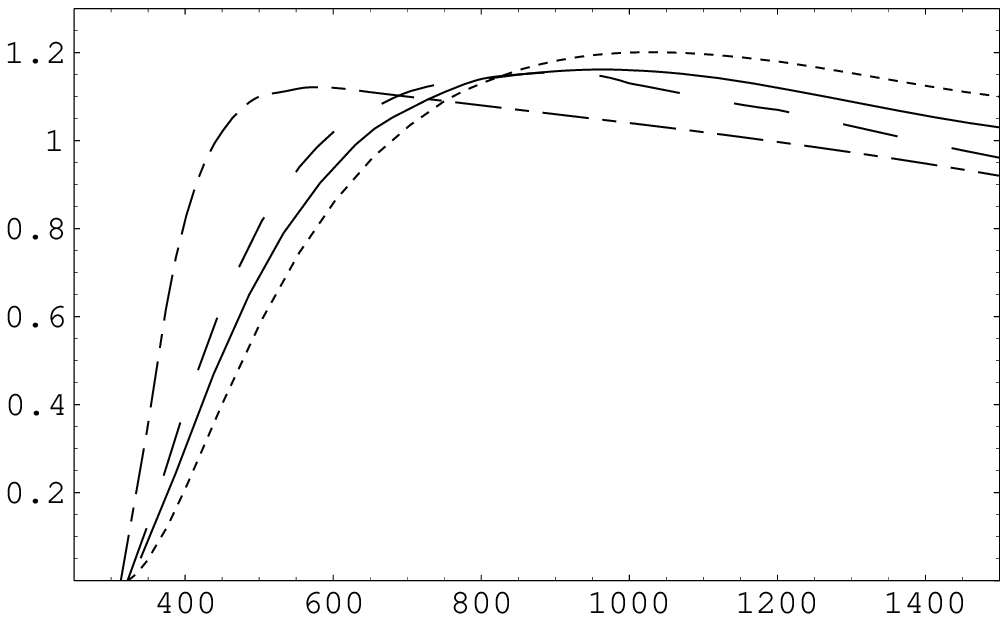}}

\put(10.3,0.1){$\sqrt{s_{ee}}/$GeV}
\put(0.1,6.4){$\sigma_p$/pb}
\end{picture}
\caption[]{\label{p1}Convoluted cross sections for 
  $\gamma\gamma\to\tilde{\chi}^+_k\tilde{\chi}^-_k$ for $m_k=128$ GeV
  as a function of the ee-CMS energy for polarizations:\\
$(\lambda_{c_1},\lambda_{L_1})=(0,0)$, $(\lambda_{c_2},\lambda_{L_2})=(0,0)$ 
(solid line)\\
$(\lambda_{c_1},\lambda_{L_1})=(0.85,0)$,
$(\lambda_{c_2},\lambda_{L_2})=(-0.85,0)$ (dashed line)\\
$(\lambda_{c_1},\lambda_{L_1})=(0.85,0)$,
$(\lambda_{c_2},\lambda_{L_2})=(0.85,0)$  (long-dashed line)\\
$(\lambda_{c_1},\lambda_{L_1})=(0.85,-1)$, 
$(\lambda_{c_2},\lambda_{L_2})=(0.85,-1)$ (dot-dashed line)}
\end{figure}

\begin{figure}[p]
\centering
\setlength{\unitlength}{1cm}

\begin{picture}(11.7,6.7)

\put(0.,-2.){\includegraphics{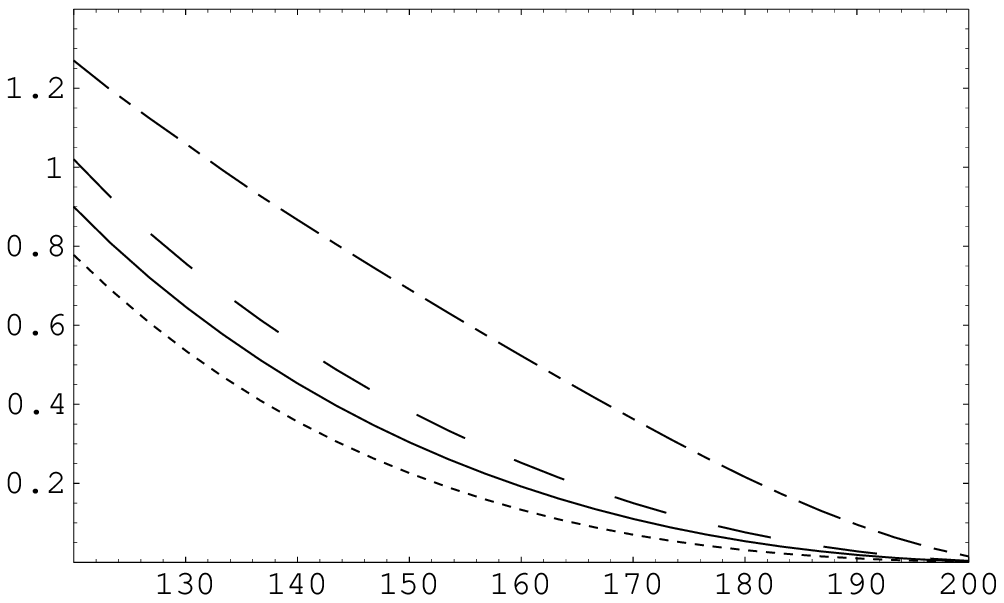}}

\put(10.4,.1){$m_k$/GeV}
\put(0.1,6.4){$\sigma_p$/pb}

\end{picture}
\caption[]{\label{p25}Convoluted cross sections for
  $\gamma\gamma\to\tilde{\chi}^+_k\tilde{\chi}^-_k$ at
  $\sqrt{s_{ee}}=500$ GeV, as a function of the chargino mass $m_k$
  for polarizations:\\
$(\lambda_{c_1},\lambda_{L_1})=(0,0)$, $(\lambda_{c_2},\lambda_{L_2})=(0,0)$ 
(solid line)\\
$(\lambda_{c_1},\lambda_{L_1})=(0.85,0)$,
$(\lambda_{c_2},\lambda_{L_2})=(-0.85,0)$ (dashed line)\\
$(\lambda_{c_1},\lambda_{L_1})=(0.85,0)$,
$(\lambda_{c_2},\lambda_{L_2})=(0.85,0)$  (long-dashed line)\\
$(\lambda_{c_1},\lambda_{L_1})=(0.85,-1)$, 
$(\lambda_{c_2},\lambda_{L_2})=(0.85,-1)$ (dot-dashed line)  
} 
\end{figure}

\begin{figure}[p]
\centering
\setlength{\unitlength}{1cm}

\begin{picture}(13.2,7.5)

\put(0,0){\epsfig{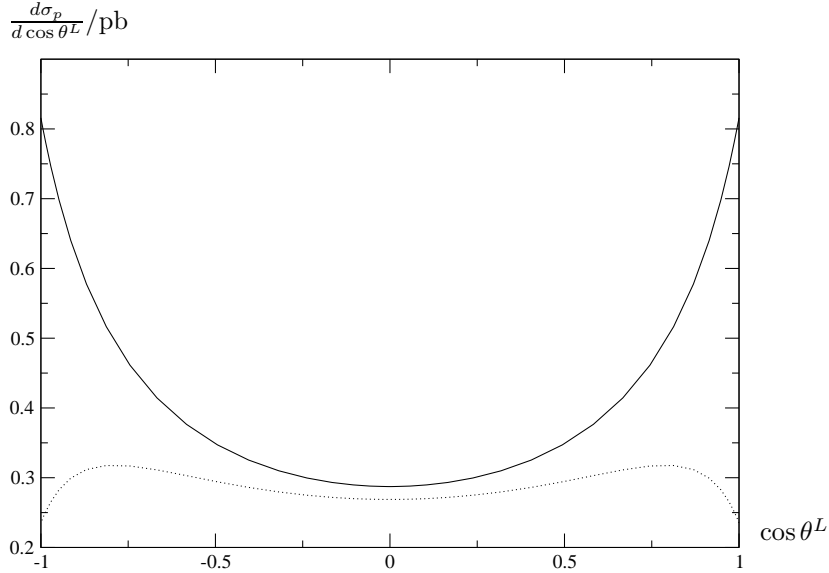}}

\put(10.,0.4){$\cos\theta^L$}
\put(0,7.2){$\frac{d\sigma_p}{d\cos\theta^L}$/pb}

\end{picture}
\caption[]{\label{p2}Angular distributions in the ee-CMS for
  $\gamma\gamma\to\tilde{\chi}^+_k\tilde{\chi}^-_k$ for $m_k=128$ GeV
  and polarizations
$(\lambda_{c_1},\lambda_{L_1})=(0.85,0)$,
$(\lambda_{c_2},\lambda_{L_2})=(0.85,0)$ (solid line), 
$(\lambda_{c_1},\lambda_{L_1})=(0.85,0)$,
$(\lambda_{c_2},\lambda_{L_2})=(-0.85,0)$ (dotted line) at
$\sqrt{s_{ee}}=500$ GeV.} 
\end{figure}

The chargino production process is in leading order perturbation theory
completely model independent and only determined by the mass $m_k$ and
the charge  of the produced fermions. In Fig.\ \ref{p1} we show
the convoluted cross section $\sigma_p$ for $m_k=128$ GeV and for
polarizations $\lambda_c=0,\pm 0.85$ of the converted electrons and 
for unpolarized or right circularly polarized 
($\lambda_L=1$) laser
photons. For high energies one obtains the highest cross section for
$(\lambda_{c_1},\lambda_{L_1})=(0.85,0)$ and
$(\lambda_{c_2},\lambda_{L_2})=(-0.85,0)$ whereas for lower energies
$\sqrt{s_{ee}}\le600$ GeV
the polarizations $(\lambda_{c_1},\lambda_{L_1})=(0.85,-1)$,
$(\lambda_{c_2},\lambda_{L_2})=(0.85,-1)$ are favored. For this combination
of polarizations the dependence of $\sigma_p$ on the chargino mass
at $\sqrt{s_{ee}}=500$ GeV is shown in Fig.\ \ref{p25}.
As a pure QED process the production cross section is forward-backward symmetric.
The shape of the angular distribution depicted in Fig.\ \ref{p2} for
$\sqrt{s}=500$ GeV, however, depends on the polarization configurations. 
For $(\lambda_{c_1}, 
\lambda_{L_1})=(0.85,0)$,
$(\lambda_{c_2},\lambda_{L_2})=(0.85,0)$ the forward and backward direction
is favored, whereas for $(\lambda_{c_1},\lambda_{L_1})=(0.85,0)$, 
$(\lambda_{c_2},\lambda_{L_2})=(-0.85,0)$ the angular distribution is
nearly isotropic.

\begin{figure}[p]
\centering
\setlength{\unitlength}{1cm}

\begin{picture}(11.7,6.7)
 
\put(0.,-2.){\includegraphics{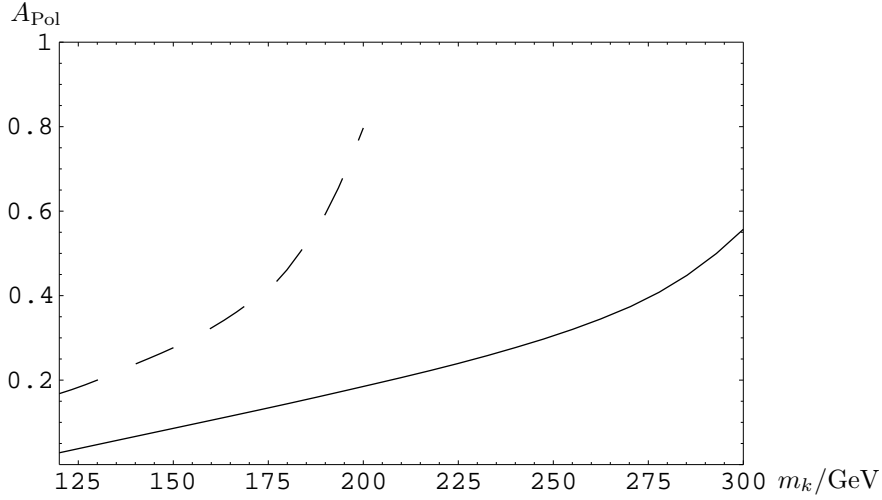}}
\put(10.3,0.1){$m_k/$GeV}
\put(0.1,6.3){$A_{\mathrm{Pol}}$}

\end{picture}
\caption[]{\label{p6}Polarization asymmetry (defined in (\ref{apol})) of the convoluted cross section
  $\gamma\gamma\to\tilde{\chi}^+_k\tilde{\chi}^-_k$
   as a function of the chargino mass $m_k$ for polarizations
$(\lambda_{c_1},\lambda_{L_1})=(0.85,1)$,
$(\lambda_{c_2},\lambda_{L_2})=(-0.85,\pm 1)$ at $\sqrt{s_{ee}}=500$ GeV
corresponding to $\sqrt{s_{\gamma\gamma}^{max}}=415$ GeV
(dashed line) and $\sqrt{s_{ee}}=800$ GeV 
corresponding to $\sqrt{s_{\gamma\gamma}^{max}}=664$ GeV
(solid line).}
\end{figure}

For polarized laser photons and electrons one can define diverse polarization
asymmetries. As an example we show in Fig.\ \ref{p6} at $\sqrt{s_{ee}}=500$ GeV
and  $\sqrt{s_{ee}}=800$ GeV the dependence on the fermion (chargino) mass 
of the polarization asymmetry 
\begin{equation}\label{apol}
A_{\mathrm{Pol}}=\frac{\sigma_p(\lambda_{c_1},\lambda_{L_1},\lambda_{c_2},\lambda_{L_2})-\sigma_p(\lambda_{c_1},\lambda_{L_1},\lambda_{c_2},-\lambda_{L_2})}{\sigma_p(\lambda_{c_1},\lambda_{L_1},\lambda_{c_2},\lambda_{L_2})+\sigma_p(\lambda_{c_1},\lambda_{L_1},\lambda_{c_2},-\lambda_{L_2})}
\end{equation}
with the polarization of only one laser pulse flipped.

\subsection{Cross section and forward-backward asymmetry of the decay leptons}

The total cross section of the decay leptons given in Table \ref{tab2} for
$\sqrt{s_{ee}}=500$ GeV and $\sqrt{s_{ee}}=800$ GeV and polarizations
$(\lambda_{c_1},\lambda_{L_1})=(0.85,0)$,
$(\lambda_{c_2},\lambda_{L_2})=(-0.85,0)$ is not affected by
spin correlations and factorizes into the chargino production cross
section and the leptonic branching ratio.  
Since, however, the angular distribution of the decay products is sensitive to
the polarization of the parent particle \cite{gudichar}, 
the forward-backward asymmetry
\begin{equation}\label{asym}
A_{\mathrm{FB}}=\frac{\sigma_e(\cos\theta_{e^+} >
  0)-\sigma_e(\cos\theta_{e^+} < 
  0)}{\sigma_e(\cos\theta_{e^+} > 0)+\sigma_e(\cos\theta_{e^+} < 0)}
\end{equation}
of the positrons from the decay
$\tilde{\chi}_1^+\to\tilde{\chi}_1^0e^+\nu_e$ may be quite
large. 
Since the production process and the transverse
polarization of the charginos $\Sigma^1_{P,\alpha\beta}$ (\ref{pol1}) is
forward-backward symmetric  $A_{\mathrm{FB}}$ will be
largest for forward-backward antisymmetric longitudinal polarization 
$\Sigma^3_{P,\alpha\beta}$. For monochromatic photons
this is the case for opposite circular polarization
($\alpha=-\beta=\pm 1$) of both beams, whereas for unpolarized photon beams or 
beams with the same polarization $\alpha=\beta$ corresponding to
$(\lambda_{c_1},\lambda_{L_1})=(\lambda_{c_2}\lambda_{L_2})$,
the longitudinal polarization $\Sigma^3_{P,\alpha\beta}$ 
is forward-backward symmetric and the forward-backward asymmetry 
$A_{\mathrm{FB}}=0$ vanishes. Analogously, the case of Compton
backscattering of laser pulses off oppositely polarized electron beams
($\lambda_{c_1}=-\lambda_{c_2}=0.85$) and unpolarized laser photons
($\lambda_{L_1}=\lambda_{L_2}=0$) results in the largest
forward-backward asymmetry of the decay leptons since for this combination
the high energy photons are polarized with mean
helicities $\lambda(y_1)>0$ and $\lambda(y_2)<0$
\cite{ginzburg}. Near threshold the spin
correlations between production and decay are strongest at the 
expense of the cross section, which is largest for the combination 
$\lambda_{c_1}=\lambda_{c_2}=0.85$ and
$\lambda_{L_1}=\lambda_{L_2}=-1$ (Fig.\ \ref{p1}). 

\begin{figure}[b]
\centering
\setlength{\unitlength}{1cm}

\begin{picture}(13.2,7.2)

\put(0,0){\epsfig{file=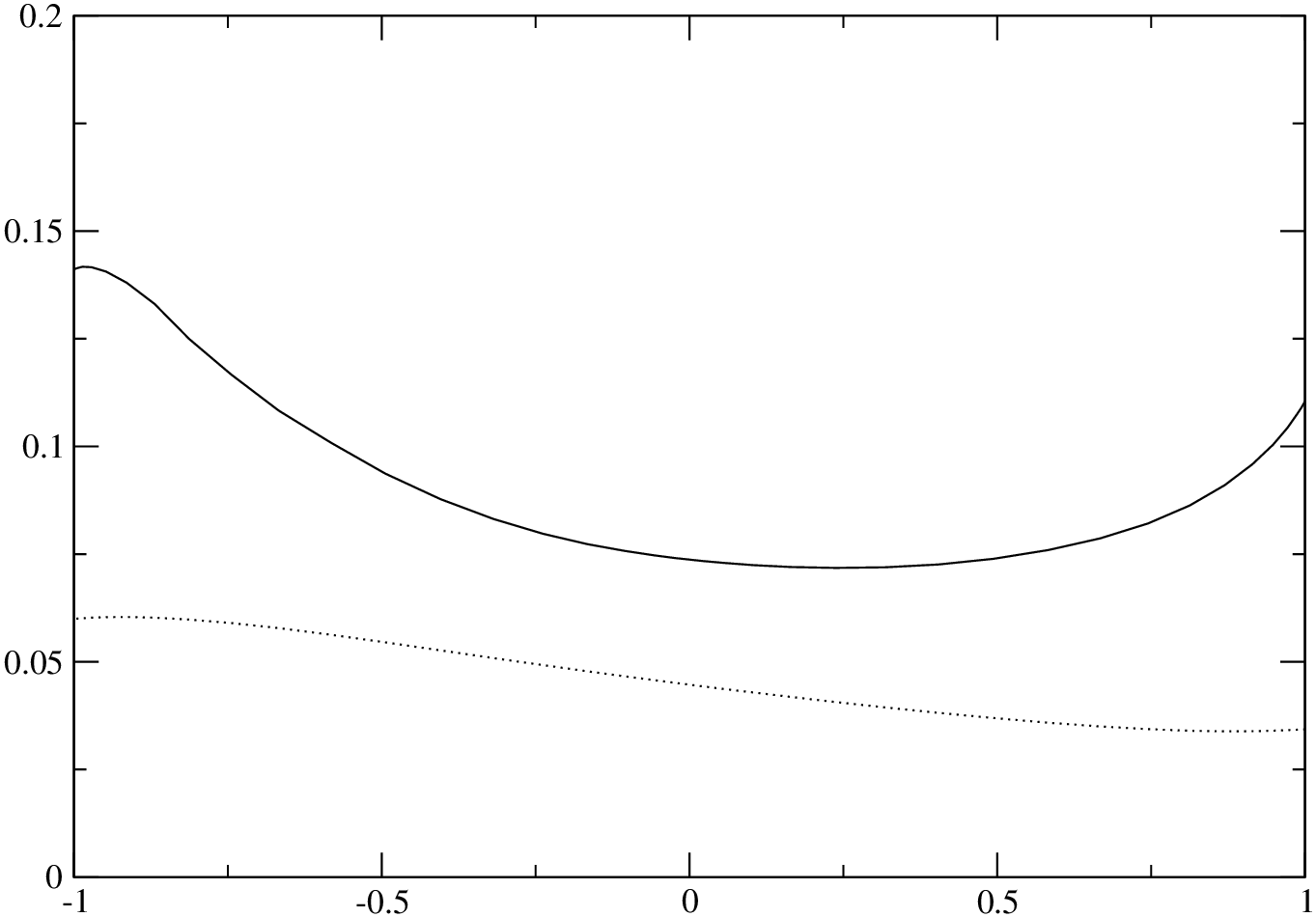,scale=0.4}}

\put(9.4,0.4){$\cos\theta_{e^+}^L$}
\put(0,7.){$\frac{d\sigma_e}{d\cos\theta_{e^+}^L}$/pb}

\end{picture}
\caption[]{\label{p4}Angular distribution in the ee-CMS of the decay positron in
  $\gamma\gamma\to\tilde{\chi}^+_1\tilde{\chi}^-_1$,
  $\tilde{\chi}^+_1\to\tilde{\chi}^0_1e^+\nu_e$
   for scenario A and polarizations
$(\lambda_{c_1},\lambda_{L_1})=(0.85,0)$,
$(\lambda_{c_2},\lambda_{L_2})=(-0.85,0)$ at $\sqrt{s_{ee}}=500$ GeV
(dotted line) and $\sqrt{s_{ee}}=800$ GeV (solid line).} 
\end{figure}

\begin{figure}[p]
\centering
\setlength{\unitlength}{1cm}

\begin{picture}(12.2,7.2)

\put(0,0){\epsfig{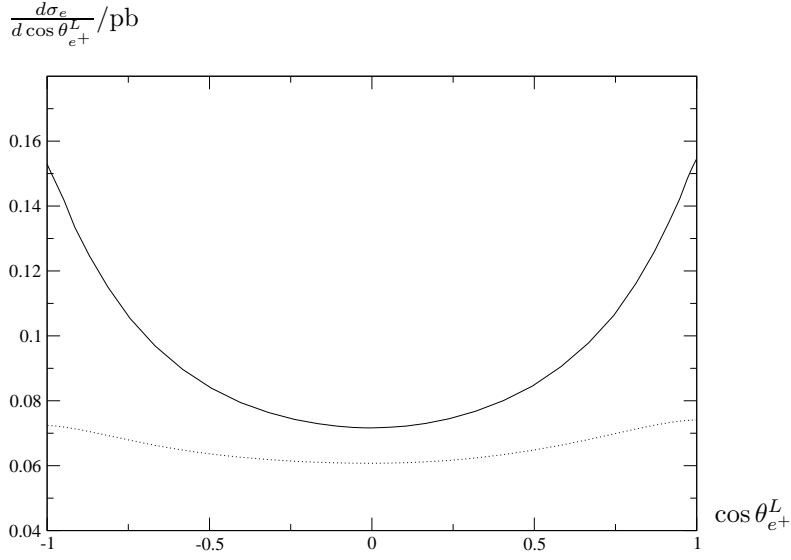}}

\put(9.4,0.4){$\cos\theta_{e^+}^L$}
\put(0,7.){$\frac{d\sigma_e}{d\cos\theta_{e^+}^L}$/pb}

\end{picture}
\caption[]{\label{p3}Angular distribution in the ee-CMS of the decay positron in
  $\gamma\gamma\to\tilde{\chi}^+_1\tilde{\chi}^-_1$,
  $\tilde{\chi}^+_1\to\tilde{\chi}^0_1e^+\nu_e$
   for scenario B and polarizations
$(\lambda_{c_1},\lambda_{L_1})=(0.85,0)$,
$(\lambda_{c_2},\lambda_{L_2})=(-0.85,0)$ at $\sqrt{s_{ee}}=500$ GeV
(dotted line) and $\sqrt{s_{ee}}=800$ GeV (solid line).} 
\end{figure}

In Figs.\ \ref{p4} and \ref{p3} we show the angular distributions for
$\lambda_{c_1}=-\lambda_{c_2}=0.85$, $\lambda_{L_1}=\lambda_{L_2}=0$
at $\sqrt{s_{ee}}=500$ GeV and $\sqrt{s_{ee}}=800$ GeV in scenario A
and scenario B. The lepton angular
distribution sensitively depends on the mixing character of the
chargino and the neutralino. In the gaugino-like scenario A it
exhibits large asymmetries, $A_{\mathrm{FB}}=-18\%$ for
$\sqrt{s_{ee}}=500$ GeV and $A_{\mathrm{FB}}=-12\%$ for
$\sqrt{s_{ee}}=800$ GeV, whereas in the higgsino-like scenario B the
angular distribution is nearly symmetric with
$A_{\mathrm{FB}}=0.83\%$ for $\sqrt{s_{ee}}=500$ GeV and
$A_{\mathrm{FB}}=0.36\%$ for $\sqrt{s_{ee}}=800$ GeV (Table~\ref{tab2}).

Since the total cross section $\sigma_e$ factorizes into the production
cross section $\sigma_p$ and the branching ratio of the chargino decay,
the polarization asymmetries of $\sigma_e$ are identical to those of
$\sigma_p$ and independent of the chargino decay.

\subsection{$M_1$-Dependence} \label{secm1}

Several studies analyze the determination of the 
gaugino mass parameter $M_1$ via production and subsequent leptonic decay 
of neutralinos or selectrons in $e^+e^-$ annihilation 
\cite{kal} --\cite{bloechi2}, via
selectron production in $e^-e^-$ scattering \cite{bloechi2, feng} or 
via associated
selectron neutralino (LSP) production in $e\gamma$ scattering 
processes \cite{bloechi1, BF}.
In all these processes the $M_1$ dependence of the cross sections 
of the decay leptons,
their polarization and forward-backward asymmetries result from a 
complex interplay
of the $M_1$ dependence of the respective production and decay mechanism.

In the process investigated here only the decay of the charginos contributes to the 
$M_1$ dependence of the cross section $\sigma_e$ and to the forward-backward
asymmetry $A_{FB}$ of the decay leptons. In the following we study the $M_1$ dependence of $\sigma_e$ 
and $A_{FB}$ fixing the other parameters as in scenario A and B.

\begin{figure}[p]
\centering
\setlength{\unitlength}{1cm}

\begin{picture}(14.,7.4)

\put(0,0){\epsfig{file=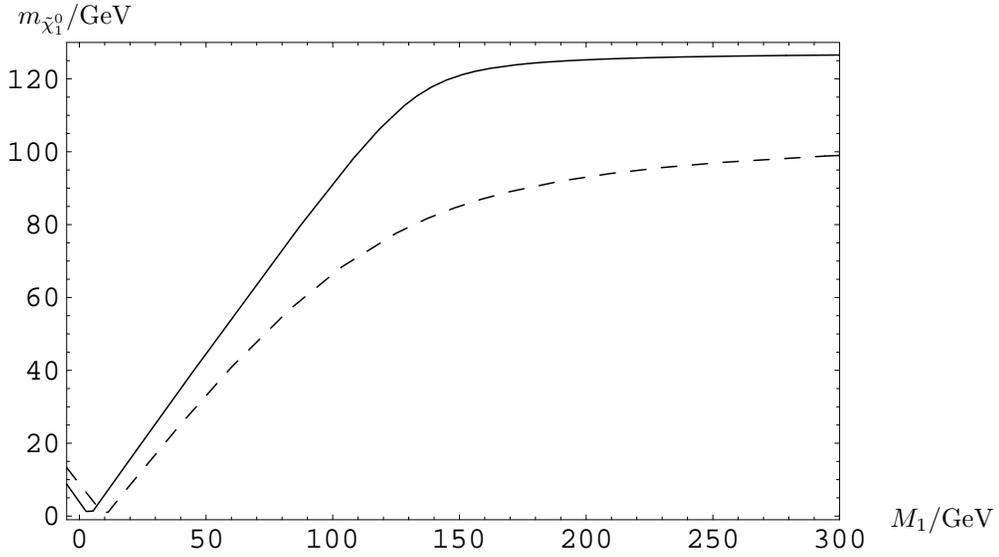,scale=1.13}}

\put(11.8,0.4){$M_1/$GeV}
\put(0.2,7.1){$m_{\tilde{\chi}_1^0}/$GeV}

\end{picture}
\caption[]{\label{p1a}$M_1$-dependence of the LSP mass for $M_2=152$
  GeV, $\mu=316$ GeV, $\tan\beta=3$
  (solid line) and for $M_2=370$
  GeV, $\mu=125$ GeV, $\tan\beta=3$ (dashed line). }
\end{figure}

\begin{figure}[p]
\centering
\setlength{\unitlength}{1cm}

\begin{picture}(11.7,6.7)

\put(0.,-2.){\includegraphics{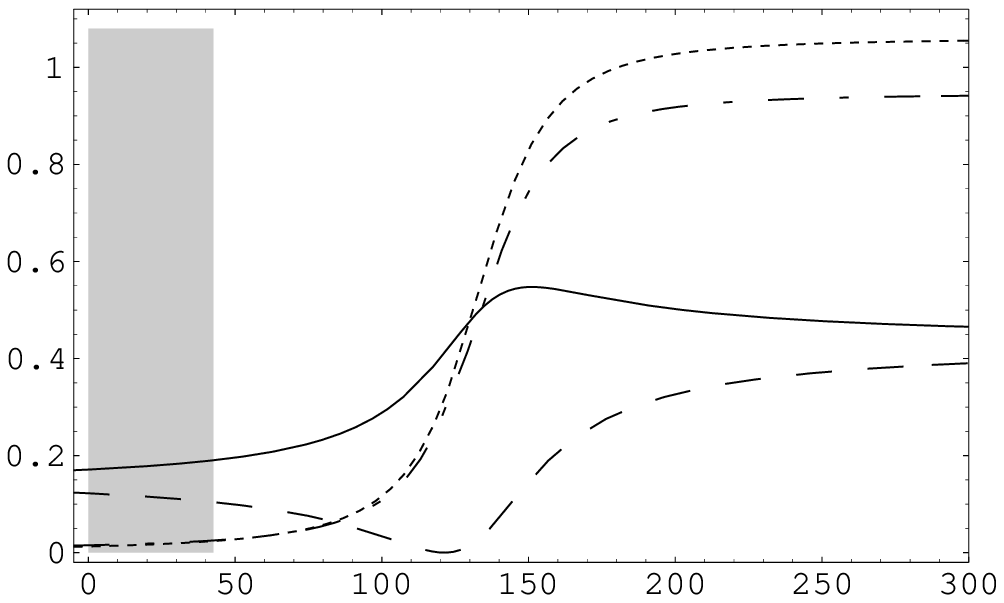}}

\put(0.1,6.35){$(f^L_{e,\nu1})^2$, $(O^{L,R}_{11})^2$} 
\put(10.3,0.1){$M_1/$GeV}

\end{picture}
\caption[]{\label{p2a}$M_1$-dependence of the couplings
  $(f^L_{\nu1})^2$ (solid line), $(f^L_{e1})^2$ (dashed line),
  $(O^{L}_{11})^2$ (long-dashed line) and
  $(O^R_{11})^2$  (dot-dashed line) as defined in
  \cite{gudichar} for $M_2=152$
  GeV, $\mu=316$ GeV, $\tan\beta=3$ (corresponding to scenario B). The shadowed region 
  is excluded by the lower bound $m_{\tilde{\chi}_1^0}>38$ GeV.}
\end{figure}

\begin{figure}[p]
\centering
\setlength{\unitlength}{1cm}

\begin{picture}(11.7,6.7)

\put(0.,-2.){\includegraphics{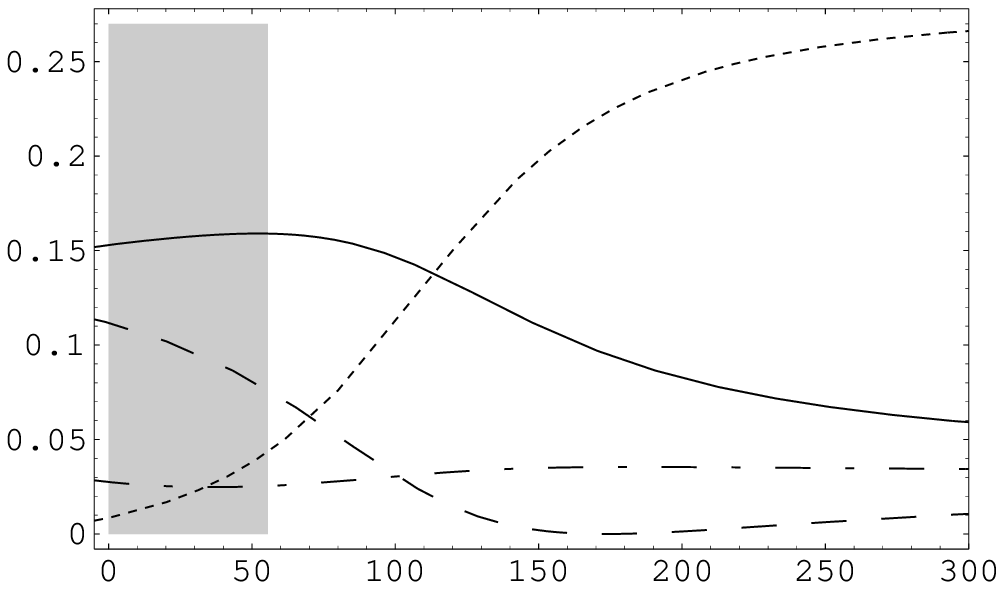}}

\put(0.1,6.35){$(f^L_{e,\nu1})^2$, $(O^{L,R}_{11})^2$}
\put(10.3,0.1){$M_1/$GeV}

\end{picture}
\caption[]{\label{p3a}$M_1$-dependence of the couplings
$(f^L_{\nu1})^2$ (solid line), $(f^L_{e1})^2$ (dashed line),
  $(O^{L}_{11})^2$ (long-dashed line) and
  $(O^R_{11})^2$ (dot-dashed line)
   as defined in
  \cite{gudichar} for $M_2=370$ GeV, $\mu=125$ GeV, $\tan\beta=3$ (corresponding to scenario B). The
  shadowed region  
  is excluded by the lower bound $m_{\tilde{\chi}_1^0}>38$ GeV.}
\end{figure}

The decay observables are affected by the $M_1$ dependence of both
the LSP mass 
$m_{\tilde{\chi}_1^0}$ and the relevant couplings displayed in 
Figs.~\ref{p1a} -- \ref{p3a}.  
The $M_1$
dependence of $m_{\tilde{\chi}_1^0}$ is very similar in both 
scenarios and shows a strong variation for $M_1<150$ GeV. The variation 
of the couplings with $M_1$, however, sensitively depends on the mixing character of 
the chargino $\tilde{\chi}_1^{\pm}$ and is in both scenarios very pronounced 
for $M_1<170$ GeV.

\begin{figure}[p]
\setlength{\unitlength}{1cm}
\begin{picture}(13,7.4)

\put(0,0){\epsfig{file=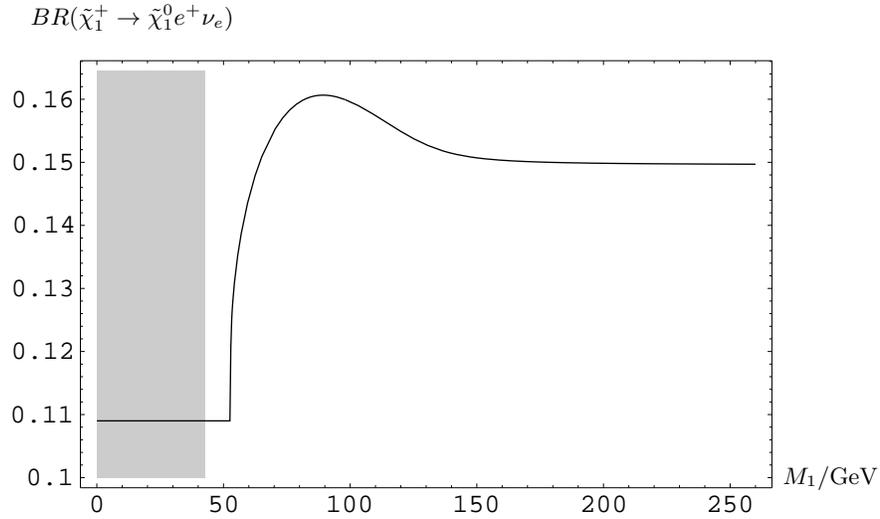,scale=1.}}
\put(0.3,6.6){\small$ BR(\tilde{\chi}^+_1\to\tilde{\chi}^0_1e^+\nu_e)$}
\put(10.3,0.5){\small$M_1/\mathrm{GeV}$}
\end{picture}

\caption[]{\label{p7}Branching ratio for 
  $\tilde{\chi}^+_1\to\tilde{\chi}^0_1e^+\nu_e$ as a function of the
  parameter $M_1$ for $M_2=152$
  GeV, $\mu=316$ GeV, $\tan\beta=3$. The shadowed region 
  is excluded by the lower bound $m_{\tilde{\chi}_1^0}>38$ GeV.}
\end{figure}

Since the $M_1$ dependence of $\sigma_e$ is exclusively
determined by the decay we show in Fig.\ \ref{p7}
the branching ratio for the decay channel
$\tilde{\chi}^+_1\to\tilde{\chi}^0_1e^+\nu_e$ for $M_2=152$ GeV,
$\mu=316$ GeV corresponding to the gaugino-like scenario A.
For $M_1<50$ GeV the two body-decay
$\tilde{\chi}^+_1\to\tilde{\chi}^0_1W^+$ is kinematically allowed and the leptonic
branching ratio of the chargino
$BR(\tilde{\chi}^+_1\to\tilde{\chi}^0_1e^+\nu_e)\approx
BR(\tilde{\chi}^+_1\rightarrow
\tilde{\chi}^0_1W^+)\times BR(W^+\rightarrow e^+\nu_e)$ is
nearly independent of $M_1$.
Between $M_1=50$ GeV and $M_1=100$ GeV the
branching ratio and consequently the cross section
$\sigma_e=\sigma_p\times
BR(\tilde{\chi}^+_1\to\tilde{\chi}^0_1e^+\nu_e)$ varies by a 
factor 1.5. 

\begin{figure}[b]
\setlength{\unitlength}{1cm}
\begin{picture}(13,7.)

\put(0,0){\epsfig{file=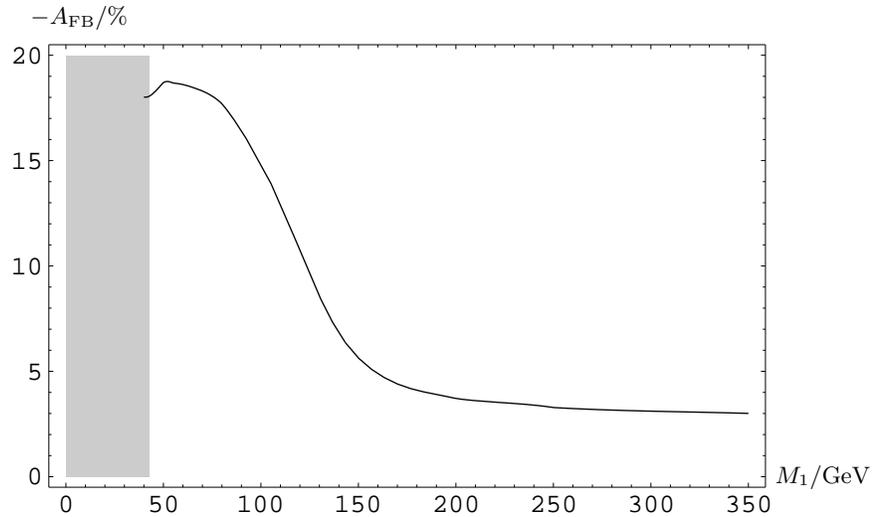,scale=1.}}
\put(0.3,6.6){\small$-A_{\mathrm{FB}}$/\%}
\put(10.2,0.5){\small$M_1/\mathrm{GeV}$}
\end{picture}

\caption[]{\label{p8}Forward-backward asymmetry in the ee-CMS of the decay positron from
  $\gamma\gamma\to\tilde{\chi}^+_1\tilde{\chi}^-_1$,
  $\tilde{\chi}^+_1\to\tilde{\chi}^0_1e^+\nu_e$ as a function of the
  parameter $M_1$ at $\sqrt{s_{ee}}=500$ GeV for
  $(\lambda_{c_1},\lambda_{L_1})=(0.85,0)$,
$(\lambda_{c_2},\lambda_{L_2})=(-0.85,0)$ for $M_2=152$
  GeV, $\mu=316$ GeV, $\tan\beta=3$. The shadowed region 
  is excluded by the lower bound $m_{\tilde{\chi}_1^0}>38$ GeV. }
\end{figure}

Provided that the parameter $M_2$ is measured in 
chargino production in $e^+e^-$ annihilation the $\gamma\gamma$ mode of a
Linear Collider provides a test of
the GUT relation betweeen $M_1$ and $M_2$. The forward-backward
asymmetry of the decay positron (Fig.\ \ref{p8}), however, turns out to
be much more sensitive on $M_1$ than the cross section $\sigma_e$.  
It changes monotonously from
19\% to 5\% in
the region 50 GeV$<M_1<150$ GeV. 

For $M_2=370$ GeV, $\mu=125$ GeV corresponding to the higgsino-like
scenario B the branching ratio and the cross section $\sigma_e$ is nearly independent of $M_1$.
In this scenario the forward-backward asymmetry is smaller than 1\% for all values of
$M_1$.

\subsection{Sneutrino mass dependence}
\label{secsneu}
In \cite{gudi2} methods have been proposed to determine the sneutrino mass from
chargino pair production and decay in $e^+e^-$-annihilation.
If the chargino has a large gaugino component sneutrino and
selectron exchange strongly influence the leptonic branching
ratio and the lepton angular distribution. In this
section we study the $m_{\tilde{\nu}_e}$ dependence of the cross section
$\sigma_e$ and the forward-backward asymmetry of the decay lepton assuming
the $SU(2)_L$ relation
(\ref{su2}) 
for $m_{\tilde{e}_L}$. 
All other parameters are fixed as in scenario A.

\begin{figure}[p]
\setlength{\unitlength}{1cm}
\begin{picture}(12.2,7.2)

\put(0,0){\epsfig{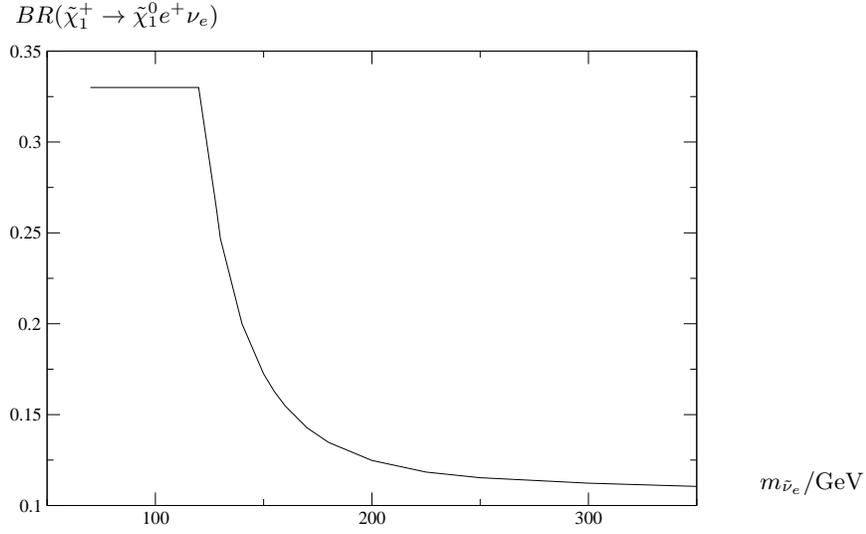}}

\put(0.1,6.7){\small$ BR(\tilde{\chi}^+_1\to\tilde{\chi}^0_1e^+\nu_e)$}
\put(10,0.5){\small$m_{\tilde{\nu}_e}/\mathrm{GeV}$}
\end{picture}

\caption[]{\label{p9}Branching ratio for 
  $\tilde{\chi}^+_1\to\tilde{\chi}^0_1e^+\nu_e$ as a function of the
  sneutrino mass $m_{\tilde{\nu}_e}$ for $M_2=152$
  GeV, $\mu=316$ GeV, $\tan\beta=3$.}
\end{figure}

\begin{figure}[b]
\setlength{\unitlength}{1cm}
\begin{picture}(13,7.)

\put(0,0){\epsfig{file=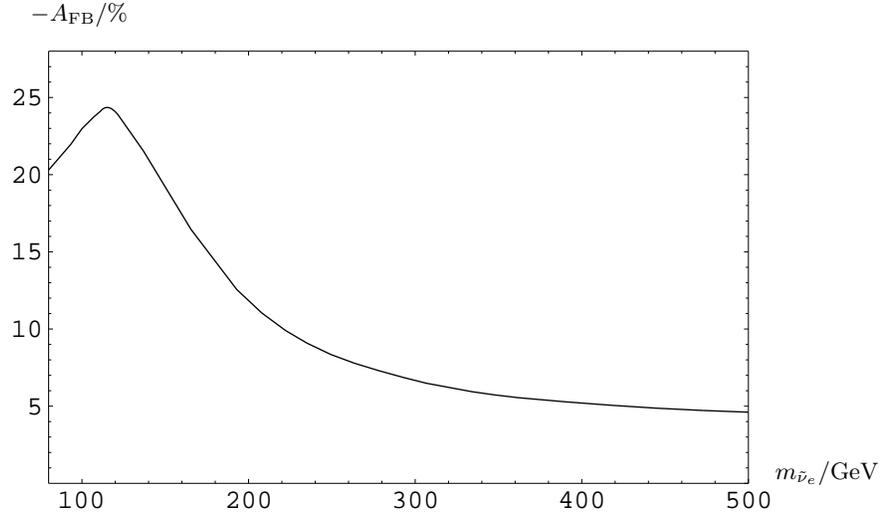,scale=1.}}
\put(0.3,6.6){\small$-A_{\mathrm{FB}}$/\%}
\put(10.2,0.5){\small$m_{\tilde{\nu}_e}/\mathrm{GeV}$}
\end{picture}

\caption[]{\label{p10}Forward-backward asymmetry in the ee-CMS of the decay positron from
  $\gamma\gamma\to\tilde{\chi}^+_1\tilde{\chi}^-_1$,
  $\tilde{\chi}^+_1\to\tilde{\chi}^0_1e^+\nu_e$ as a function of the
  sneutrino mass $m_{\tilde{\nu}_e}$ at $\sqrt{s_{ee}}=500$ GeV for
  $(\lambda_{c_1},\lambda_{L_1})=(0.85,0)$,
$(\lambda_{c_2},\lambda_{L_2})=(-0.85,0)$ for $M_2=152$
  GeV, $\mu=316$ GeV, $\tan\beta=3$.}
\end{figure}

As a function of the sneutrino mass we show in Fig.~\ref{p9} the branching ratio
$BR(\tilde{\chi}^+_1\rightarrow\tilde{\chi}^0_1e^+\nu_e)$ and
in Fig.~\ref{p10} the forward-backward asymmetry 
for $\sqrt{s_{ee}}=500$ GeV and beam
polarizations $(\lambda_{c_1},\lambda_{L_1})=(0.85,0)$ and
$(\lambda_{c_2},\lambda_{L_2})=(-0.85,0)$.
These polarization configurations result in the largest
forward-backward asymmetry.
The cross section shows a pronounced $m_{\tilde{\nu}_e}$ dependence for
$m_{\tilde{\nu}_e}\lessapprox 250$ GeV whereas
the forward-backward asymmetry exhibits an appreciable
$m_{\tilde{\nu}_e}$ dependence up to $m_{\tilde{\nu}_e}\sim 400$
GeV, which is considerably beyond the kinematical limit for 
sneutrino pair production in $e^+e^-$-annihilation at $\sqrt{s_{ee}}=500$
GeV. 

For $m_{\tilde{\nu}_e}<m_{\tilde{\chi}^+_1}$  the 
two-body decays $\tilde{\chi}^+_1\rightarrow l^+\tilde{\nu}_l$, $l=(e,\mu)$
and eventually $l=(e,\mu,\tau)$ are dominating. Since 
for our set of parameters the sneutrino decays completely invisible via
$\tilde{\nu}_l\rightarrow\tilde{\chi}^0_1\nu_l$ the branching ratio for
$\tilde{\chi}^+_1\rightarrow\tilde{\chi}^0_1e^+\nu_e$ is given by that for
the two-body decay into positron and sneutrino and therefore
independent of the sneutrino mass.

For $m_{\tilde{\nu}_e}>m_{\tilde{\chi}^+_1}$ , however, 
the $m_{\tilde{\nu}_e}$ dependence of both the cross
section and the forward-backward asymmetry is free of any
ambiguities. With increasing sneutrino mass the contributions
from $\tilde{\nu}_e$ and $\tilde{e}_L$ exchange are more and more
suppressed so that finally only the contribution from $W$-exchange
survives. We conclude that measuring the forward-backward asymmetry
with suitably polarized beams is a useful method for the determination of the
sneutrino mass. For a quantitative evaluation of the accuracy Monte
Carlo studies would be necessary.

\section{Conclusion}

Pair production of charginos with subsequent decay in photon-photon
collisions allows to study the decay
mechanism separately from production. 
We have presented analytical formulae for the polarization and the
spin-spin correlations of fermions produced in collisions of circularly
polarized photon beams.

For high energy photons from
Compton backscattering of polarized laser pulses off polarized electron
beams we calculated the production cross section of the lighter chargino
and the cross section, the angular distribution and the forward-backward
asymmetry of the positron from the leptonic $\tilde{\chi}^+_1$
decay. We have shown that for gaugino-like chargino and LSP  
the cross section and particularly the forward-backward asymmetry of the decay leptons
is sensitive to the gaugino mass parameter $M_1$ and to the sneutrino mass
and allows to constrain them. Contrary to chargino production 
in electron-positron annihilation 
neither the dependence on
the gaugino mass parameter $M_1$ nor
the dependence on the sneutrino mass of the cross section and the
forward-backward asymmetry show ambiguities
above the threshold
for two-particle decay of the chargino. 

\section*{Acknowledgements}
This work was supported by the Deutsche Forschungsgemeinschaft, 
contract FR 1064/4-1 and the Bundesministerium 
f\"ur Bildung und Forschung, contract 05 HT9WWa 9.

\begin{table}[p]
\centering
\renewcommand{\arraystretch}{1.5}
\begin{tabular}{||c||c|c||} \hline\hline
  \multicolumn{3}{||c||}{Scenario A} \\ \hline
  \multicolumn{3}{||c||}{$M_1=76$ GeV \hspace{5ex}  $M_2=152$ GeV
    \hspace{5ex} $\mu=316$ GeV} \\
  \multicolumn{3}{||c||}
    {\hspace{2.9ex}$\tan\beta=3$ \hspace{7.2ex}
    $m_{\tilde{\nu}}=158$  GeV \hspace{3.3ex}   $m_0=100$ GeV} \\ \hline\hline 
  & $\tilde{\chi}^+_1$ & $\tilde{\chi}^0_1$ \\ \hline
  mass / GeV & $128$ & $69.5$ \\ \hline
  mixing & \begin{minipage}{3.1cm}\begin{center}$(\tilde{W}^+|\tilde{H}^+)$\\$(0.96|-0.28)$\end{center} \end{minipage}&\begin{minipage}{5.1cm}\begin{center}$(\tilde{\gamma}|\tilde{Z}|\tilde{H}_a|\tilde{H}_b)$\\$(0.78|-0.59|0.15|0.14)$\end{center}\end{minipage} \\ \hline \hline
 \multicolumn{3}{||c||}{Scenario B} \\ \hline
  \multicolumn{3}{||c||}{$M_1=185$ GeV \hspace{4.4ex}  $M_2=370$ GeV
    \hspace{5ex} $\mu=125$ GeV} \\
  \multicolumn{3}{||c||}
    {\hspace{2.9ex}$\tan\beta=3$ \hspace{7.2ex}
    $m_{\tilde{\nu}}=339$  GeV \hspace{3.3ex}   $m_0=100$ GeV} \\ \hline\hline 
  & $\tilde{\chi}^+_1$ & $\tilde{\chi}^0_1$ \\ \hline
  mass / GeV & $108$ & $91.6$ \\ \hline
  mixing & \begin{minipage}{3.1cm}\begin{center}$(\tilde{W}^+|\tilde{H}^+)$\\$(0.33|-0.94)$\end{center} \end{minipage}&\begin{minipage}{5.1cm}\begin{center}$(\tilde{\gamma}|\tilde{Z}|\tilde{H}_a|\tilde{H}_b)$\\$(-0.21|0.37|-0.78|-0.47)$\end{center} \end{minipage} \\ \hline \hline
 
\end{tabular}
\caption[]{\label{szen}Masses and mixing character of the light chargino
  $\tilde{\chi}_1^\pm$ and the lightest neutralino
  $\tilde{\chi}_1^0$ in scenarios A and B.} 
\end{table}

\begin{table} \centering \renewcommand{\arraystretch}{1.4}
\begin{tabular}{|c|c|c|c|c|}
\hline
 & \multicolumn{2}{c|}{Scenario A} & \multicolumn{2}{c|}{Scenario B} \\ \hline
\hline $\sqrt{s_{ee}}$/GeV & 500 & 800 & 500 & 800 \\ \hline \hline
$\sigma_e$/pb & 0.09 & 0.18 & 0.13 & 0.19 \\ \hline
$A_{\mathrm{FB}}$/\% & -18 & -12 & 0.83 & 0.36 \\
\hline
\end{tabular}
\caption[]{\label{tab2}Convoluted cross sections and positron forward-backward asymmetries 
for $\gamma\gamma\to \tilde{\chi}^+_1 \tilde{\chi}^-_1$,
$\tilde{\chi}^+_1\to\tilde{\chi}^0_1 e^+\nu_e$
in scenario A and B at $\sqrt{s_{ee}}=500$ GeV and
$\sqrt{s_{ee}}=800$ GeV for polarizations
$(\lambda_{c_1},\lambda_{L_1})=(0.85,0)$,  
$(\lambda_{c_2},\lambda_{L_2})=(-0.85,0)$.}
\end{table}

\end{document}